\documentclass[12pt]{article}
\usepackage{latexsym}
\usepackage{graphicx}

\thicklines                         
\long \def \blockcomment #1\endcomment{}

\begin{document}           
\baselineskip=0.33333in

\begin{quote} \raggedleft TAUP 2871-2007
\end{quote}
\vglue 0.5in
\begin{center}{\bf Mathematical Constraints on Gauge \\
in Maxwellian Electrodynamics}
\end{center}
\begin{center}E. Comay$^*$
\end{center}

\begin{center}
School of Physics and Astronomy \\
Raymond and Beverly Sackler Faculty of Exact Sciences \\
Tel Aviv University \\
Tel Aviv 69978 \\
Israel
\end{center}
\vglue 0.5in
\noindent
\vglue 0.2in
\noindent
Abstract:

The structure of classical
electrodynamics based on the variational principle
together with causality and space-time homogeneity is analyzed. It
is proved that in this case the 4-potentials are defined uniquely.
On the other hand, the approach where Maxwell equations and the Lorentz
law of force are regarded as cornerstones of the theory allows gauge
transformations. For this reason, the two theories are
{\em not equivalent}.
A simple example substantiates this conclusion.
Quantum physics is linked to the variational principle and it is
proved that the same result holds for it. The compatibility of this
conclusion with gauge invariance of the Lagrangian density is explained.
Several alternative possibilities that may follow this work are pointed out.

\newpage
\noindent
{\bf 1. Introduction}
\vglue 0.33333in

One may regard the equations of motion of a physical system as the
fundamental elements of a theory. Thus, the equations of motion can
be used for deriving useful formulas that describe properties of
the system. However, it is now recognized that other principles
play a more profound role. Using this approach, the variational
principle, causality and homogeneity of space-time are regarded
here as the basis for the discussion. The present work examines
these approaches within the validity domains of classical
electrodynamics and of the associated quantum physics.
Thus, the electrodynamic theory that regards
Maxwell equations and the Lorentz law of force as cornerstones of
the theory is called here Maxwell-Lorentz electrodynamics (MLE).
The theory that relies on the variational principle is called here
variational electrodynamics (VE). MLE and VE are very closely related
theories. Thus, Maxwell
equations and the Lorentz law of force can be derived from the
variational principle (see [1], pp. 49-51,70,71,78-80;
[2], 572-578,595-597). On the other hand, MLE and VE rely on two
different sets of axioms. Therefore, the validity of their
equivalence is not a priori clear. The first part of the
discussion carried out here analyzes the two approaches
within the realm of classical electrodynamics and proves that
MLE is {\em not equivalent} to VE and that VE imposes
further restrictions on the theory's structure. Quantum mechanics is
strongly linked to the variational approach (see [3], pp. 2-23).
Thus, it is proved in this work that the same results are obtained
for quantum mechanics.

The specific subject discussed here is the role of gauge transformations
and of gauge invariance in MLE and in VE. The following argument
indicates the need for a further clarification of this subject.  It is
very well known that all terms of a physical expression must have the
same dimensions (otherwise, a change in the unit system destroys numerical
balance). Now, let $F(q)$
be an analytic function used in a description of a
physical relation. If the power series of $F(q)$
contains more than one term (e.g. $aq^m + bq^n$, where $a$ and $b$
are nonzero pure numbers and $m\ne n$),
then it is required that  $q$ be dimensionless. Thus, for example, the
exponential factor used in the Maxwell-Boltzmann distribution takes
the form $e^{-E/KT}$ and the product $KT$ has the dimensions of energy.
If $F(q)$ belongs
to a relativistic expression then covariance arguments prove that
$q$ must also be a Lorentz scalar.
The wave function's phase $e^{i({\bf k\cdot x}- \omega t)}$ satisfies
the two requirements.

   Now, let $\Phi (x^\mu )$ be a gauge function used in
VE and its 4-derivative
$\Phi (x^\mu )_{,\nu }$ is subtracted from a 4-potential in a gauge
transformation. In quantum mechanics, the charged particle's
sector contains the gauge dependent factor $e^{ie\Phi (x^\mu )}$
(see [4], p. 78).
Note that the symbol $e$ in the exponent denotes the particle's electric
charge. Now, in the system of units used here (see later in this Section)
the electric charge is a pure number $e^2\simeq 1/137$. Thus, the analytic
properties of the exponential function and the laws described in the
previous paragraph prove that in quantum mechanics, the gauge function
$\Phi (x^\mu )$ must be a dimensionless
Lorentz scalar. As of today, this restriction is not implemented
and the standard gauge transformation used in the literature regards
$\Phi (x^\mu )$ as a free function of space-time coordinates (see
[1], p. 52; [4], p. 78). This
example provides a reason for the investigation of the role
of gauge transformations which is carried out here.

   It is interesting to note that other problems emerging from gauge
transformation are already pointed out in the literature (see [2],
pp. 222, 223). Thus, in a Coulomb gauge,
a transverse electric current is found
throughout the entire space, in spite of actual charge localization.

It is proved in this work that if one adheres to VE
together with causality and space-time homogeneity
then the 4-potentials of electrodynamics
are defined uniquely. On the other hand, the 4-potentials play no
explicit role
in Maxwell equations and in the Lorentz law of force. Hence, one may
apply any gauge transformation without affecting MLE.
This is the underlying reason for the claim that
MLE is {\em not equivalent} to VE.

   In the present work, units where the speed of light
$c = 1$ and $\hbar = 1$ are used. Thus, one kind of dimension
exists and the length $[L]$ is used for this purpose.
Greek indices run from 0 to 3. The metric is diagonal and its entries are
(1,-1,-1,-1). The symbol $_{,\mu }$ denotes the partial differentiation
with respect to $x^\mu $. $A_\mu $ denotes the 4-potentials and $F^{\mu \nu }$
denotes the antisymmetric tensor of the electromagnetic fields
\begin{equation}
F^{\mu \nu } = g^{\mu \alpha}g^{\nu \beta}(A_{\beta ,\alpha} -
A_{\alpha ,\beta})
=\left(
\begin{array}{cccc}
0   & -E_x & -E_y & -E_z \\

E_x &  0   & -B_z &  B_y \\

E_y &  B_z &   0  & -B_x \\

E_z & -B_y &  B_x &  0
\end{array}
\right).
\label{eq:FMUNU}
\end{equation}

In the second Section, the main point of this work is proved for classical
physics. The third Section describes a specific example that substantiates
the proof included in Section 2. The fourth Section proves that the same
results are obtain for quantum physics.
Several implications that may be connected to the analysis presented herein
are discussed in the fifth Section.
The last Section contains concluding remarks.

\vglue 0.66666in
\noindent
{\bf 2. Gauge Transformations and Variational Electrodynamics}
\vglue 0.33333in

The standard form of the
Lagrangian density used for a derivation of Maxwell equations is
(see [1], pp. 78-80; [2], pp. 596-597)
\begin{equation}
{\mathcal L} =
- \frac {1}{16\pi }F^{\mu \nu }F_{\mu \nu } - j^\mu A_\mu ,
\label{eq:LAGR}
\end{equation}
where the first term represents free fields and the second term represents
the interaction of the fields with charged matter.
The following analysis examines a closed system of charges and fields.
For the simplicity of the discussion, let us examine the fields
associated with one charged particle $e$ whose motion is given.
This approach can be justified because, due to
the linearity of Maxwell equations, one finds that the fields of
a closed system of charges is a superposition of the fields of each
individual charge belonging to the system.
Let us examine the electromagnetic fields at a given space-time point
$x^\mu $. Using Maxwell equation
and the principle of causality, one can derive the retarded
Lienard-Wiechert 4-potentials (see [1], pp. 173-174; [2], pp. 654-656)
\begin{equation}
A_\mu = e\frac {v_\mu }{R^\alpha v_\alpha }.
\label{eq:LWPOTENTIAL}
\end{equation}
Here $v_\mu $ is the charge's 4-velocity at the retarded time and
$R^\mu $ is the 4-vector from the retarded space-time point to the
field point $x^\mu $. These 4-potentials define the fields uniquely.

A gauge transformation of $(\!\!~\ref{eq:LWPOTENTIAL})$
is (see [1], pp. 52-53; [2], pp. 220-223)
\begin{equation}
A'_\mu = A_\mu - \Phi {,_\mu} .
\label{eq:GAUGE}
\end{equation}
In the following lines, the laws of VE are used in an investigation of
the form of the gauge function $\Phi (x^\mu)$.

Relying on the variational principle, one finds constraints on
terms of the Lagrangian density. Thus, the action is a Lorentz
scalar and in the unit system used here where $\hbar=1$,
it is dimensionless. This property means that every term of the
Lagrangian density must have the dimension $[L^{-4}]$.
Now, components of the 4-current $j^\mu $
represent charge and current densities and their dimension is
$[L^{-3}]$. Therefore, the 4-potentials $A_\mu $ must be a
4-vector whose dimension is $[L^{-1}]$. These requirements are
satisfied by the Lienard-Wiechert 4-potentials $(\!\!~\ref{eq:LWPOTENTIAL})$.
Thus, also $\Phi _{,\mu }$ of $(\!\!~\ref{eq:GAUGE})$
is a 4-vector whose dimension is $[L^{-1}]$ and $\Phi $ must be a
dimensionless Lorentz scalar function of the space-time coordinates.

Now, the coordinates are entries of a 4-vector. Let us first find
the general form of a physically acceptable Lorentz scalar function
depending {\em only} on the space-time coordinates. The following
expression is a scalar function of the coordinates
\begin{equation}
f_{a,b}(x^\mu ) = (x^\mu - x_a^\mu)(x_\mu - x_{b\mu}),
\label{eq:SCALAR2}
\end{equation}
where $x_a^\mu $ and $x_b^\mu $
denote specific space-time points. The first objective is to find a
definition of the form of one scalar term $T$.
This term must be a tensorial expression which is completely contracted.
Thus, it can be cast into a
product of powers of functions like $(\!\!~\ref{eq:SCALAR2})$
\begin{equation}
T = f_{a,b}^\alpha (x^\mu ) f_{c,d}^\beta (x^\mu ) ... f_{u,v}^\gamma (x^\mu ),
\label{eq:SCALAR3}
\end{equation}
where Latin subscripts
denote specific coordinate points and Greek letters
denote the power of each function.
It follows that any scalar function of the
coordinates can be written as a sum of terms where each of which is
a product of positive or negative
powers of functions like $(\!\!~\ref{eq:SCALAR2})$.

Relying on causality and
homogeneity of space-time, one finds that in the case discussed
here there is just one specific point $x_a^\mu$, which is the
retarded position of the charge. Thus, $(\!\!~\ref{eq:SCALAR2})$
boils down into the following form
\begin{equation}
f_{a,b}(x^\mu ) \rightarrow R^\mu R_\mu .
\label{eq:R2}
\end{equation}
This outcome proves that the gauge function $\Phi (x^\mu )$,
which is a dimensionless
quantity, must be a constant. (As a matter of fact, the retardation
conditions prove that $(\!\!~\ref{eq:R2})$ vanishes identically.)

At this point it is clear that the expression
$(x^\mu - x_a^\mu)(x_{c\mu} - x_{b\mu})$ cannot be used
in place of $(\!\!~\ref{eq:SCALAR2})$. Indeed, here
$x_{c\mu} = x_{b\mu}$ and the second factor vanishes.

These arguments complete the proof showing that if one adheres to VE then
the gauge function $\Phi $
is a constant and the gauge 4-vector $\Phi _{,\mu }$ vanishes identically.
Hence, the Lienard-Wiechert 4-vector $(\!\!~\ref{eq:LWPOTENTIAL})$ is
unique.

\vglue 0.66666in
\noindent
{\bf 3. An Example}
\vglue 0.33333in

   Let us examine a simple system which consists of one motionless
particle whose mass and charge are $m$, $e$, respectively. The particle
is located in a spatial region
where the external fields vanish. Therefore, the
Lorentz force exerted on the particle vanishes too and it
remains motionless as long as these conditions do not change. Thus,
the system's energy is a constant of the motion. This property holds
for MLE, where the particle's energy is a constant
\begin{equation}
E = m.
\label{eq:EEQM}
\end{equation}

   Now, let us examine this system from the point of view of VE. For
this purpose, the external 4-potentials should be defined. Thus, the
null external fields are derived from null 4-potentials
\begin{equation}
A_{(ext)\mu } = 0\;\rightarrow \; F^{\mu \nu}_{(ext)}=0.
\label{eq:A0}
\end{equation}

   In order to define the particle's energy
one must construct the Hamiltonian. Here
the general expression is (see [1], pp. 47-49; [2], pp. 575)
\begin{equation}
H=[m^2 + ({\bf P} - e{\bf A})^2]^{1/2} + e\phi ,
\label{eq:HAM}
\end{equation}
where ${\bf P}$ denotes the canonical momentum and the components of
the 4-potentials are $(\phi,{\bf A})$. Substituting the null values of
$(\!\!~\ref{eq:A0})$ into $(\!\!~\ref{eq:HAM})$ and putting there
${\bf P} = 0$ for the motionless particle and the vanishing 4-potentials,
one equates the energy
to the Hamiltonian's value and obtains
\begin{equation}
E = m.
\label{eq:EHAMM}
\end{equation}
At this point, one finds that result $(\!\!~\ref{eq:EEQM})$ of MLE is
identical to $(\!\!~\ref{eq:EHAMM})$ of VE.

   This system is used here as an example showing how far can one
proceed if gauge transformation freedom is permissible. To
this end, let us apply a specific gauge transformation to the null external
4-potentials $(\!\!~\ref{eq:A0})$. The gauge function and its 4-potentials
are
\begin{equation}
\Phi = t^2 \;\rightarrow \; A'_{(ext)\mu} = -\Phi _{,\mu} = (-2t,0,0,0).
\label{eq:GT}
\end{equation}
In MLE nothing changes, because the equations of motion depend on
electromagnetic fields and their null value does not change
\begin{equation}
F'^{\mu \nu} = F^{\mu \nu} = 0.
\label{eq:FEQF}
\end{equation}
Hence, the energy value $(\!\!~\ref{eq:EEQM})$ continues to hold and
the gauge transformation $(\!\!~\ref{eq:GT})$ is acceptable in MLE.

   The following points show several arguments proving that
this conclusion does not hold for the VE theory.
\begin{itemize}
\item[{1.}] The gauge function of $(\!\!~\ref{eq:GT})$ has the dimensions
$[L^2]$, whereas in VE it must be dimensionless.
\item[{2.}] The gauge function of $(\!\!~\ref{eq:GT})$ is the entry $U^{00}$
of the second rank tensor $U^{\mu \nu } = x^\mu x^\nu $. On the other
hand, in VE the gauge function must be a Lorentz scalar.
\item[{3.}] Substituting the gauge 4-vector $A'_{(ext)\mu }$ of
$(\!\!~\ref{eq:GT})$ into the Hamiltonian $(\!\!~\ref{eq:HAM})$, one finds
\begin{equation}
H' = m - 2et.
\label{eq:HGT}
\end{equation}
Hence, if gauge transformations are allowed in VE then the energy of a
closed system is not a constant of the motion.
\item[{4.}] The previous argument can be observed from another point of
view. Thus, the physical state of the single motionless particle is
time independent. Hence, one expects that energy is a constant of
the motion. This point holds within the framework of
MLE which is unaffected by any gauge transformation.
It is also satisfied in VE,
provided one uses the null 4-potential $(\!\!~\ref{eq:A0})$. However,
the gauge degree of freedom allows one to use the gauge transformation
$(\!\!~\ref{eq:GT})$. This transformation casts the trivial
time-independent Hamiltonian into the time-dependent
expression $(\!\!~\ref{eq:HGT})$.
As is well known, if the Hamiltonian is time-dependent then
energy is not a constant of the motion (see [5], p. 132).
Hence, an application
of the gauge degree of freedom deprives the VE theory from having
an acceptable
expression for the energy of a physically time-independent state.
Here one finds a specific example showing that MLE and VE are not
equivalent theories.
\end{itemize}
These four conclusions prove that the gauge degree of freedom destroys VE.

\vglue 0.66666in
\noindent
{\bf 4. Gauge Transformations and Quantum Physics}
\vglue 0.33333in

As stated in the Introduction, quantum physics is very closely related
to VE. Moreover, the Ehrenfest theorem (see [6], pp. 25-27, 138) shows
that the classical limit of quantum mechanics agrees with the laws of
classical physics. For these reasons, one expects that the laws of VE
are relevant to quantum physics. A direct examination of gauge
transformations proves this matter.

The Lagrangian density of the Dirac field is (see [3], p. 84; [4],
p. 78)
\begin{equation}
{\mathcal L} = \bar \psi[\gamma ^\mu (i\partial _\mu - eA_\mu) - m]\psi ,
\label{eq:DIRACLD}
\end{equation}
This Lagrangian density yields the Dirac Hamiltonian
(see[7], p. 48)
\begin{equation}
H = \mbox {\boldmath $\alpha \,\cdot $} ({\bf P} - e{\bf A}) + \beta m +
e\phi .
\label{eq:DIRACH}
\end{equation}

Now, in quantum mechanics, the gauge transformation $(\!\!~\ref{eq:GAUGE})$
is accompanied by an appropriate transformation of the particle's wave
function. Thus, the quantum mechanical form of gauge transformation is
(see [4], p. 78)
\begin{equation}
A'_\mu = A_\mu - \Phi {,_\mu} ;\;\;\;
\psi '(x^\mu ) = e^{ie\Phi(x^\mu)}\psi (x^\mu )
\label{eq:GAUGEQM}
\end{equation}
(Note that the symbol $e$ in the exponent denotes the particle's electric
charge.)
Substituting the gauge transformation $(\!\!~\ref{eq:GAUGEQM})$
into the Lagrangian density $(\!\!~\ref{eq:DIRACLD})$, one realized
that it is gauge invariant indeed (see e.g. [4], p. 78).

It is interesting to note that in quantum mechanics the gauge function
$\Phi $ is used in the exponent of the particle's wave function. As
explained in the introduction, general laws of physics restrict
$\Phi(x^\mu)$ to be a dimensionless Lorentz scalar.
This form substantiates the classical arguments
presented in Section 2, where it is explained why VE requires that
the gauge function should be a dimensionless Lorentz scalar.

   Now let us examine the quantum mechanical version of the example
discussed in Section 3. The Dirac wave function of the spin-up
state of a motionless particle is (see [7], p. 10)
\begin{equation}
\psi (x^\mu ) = e^{-imt}(1,0,0,0).
\label{eq:PSI0}
\end{equation}
Thus, one uses the fundamental quantum mechanical equation and obtains
the particle's energy from an application of the
Dirac Hamiltonian to the wave function $(\!\!~\ref{eq:PSI0})$
\begin{equation}
E\psi = H\psi =
i\frac {\partial \psi }{\partial t}  = m\psi \rightarrow E=m.
\label{eq:EQM}
\end{equation}

Now, let us examine the gauge transformation $(\!\!~\ref{eq:GAUGEQM})$ for
the specific case $(\!\!~\ref{eq:GT})$. The
wave function $(\!\!~\ref{eq:PSI0})$ transforms as follows
\begin{equation}
\psi '(x^\mu ) = e^{iet^2}e^{-imt}(1,0,0,0).
\label{eq:PSITAG}
\end{equation}
Using the gauge transformed wave function $(\!\!~\ref{eq:PSITAG})$, one
applies a straightforward calculation and obtains the expectation value of
the Hamiltonian. Here the result differs from the original value
\begin{equation}
H'\psi '= i\frac {\partial \psi '}{\partial t}  = (m - 2et)\psi '
\rightarrow \;<H'> = m - 2et.
\label{eq:EQMTAG}
\end{equation}

This is precisely the same discrepancy which was found above for the
gauge transformation of VE of classical physics $(\!\!~\ref{eq:HGT})$.
Indeed, a gauge transformation casts a time independent Hamiltonian
into a time-dependent expression and energy calculation is destroyed.
Thus, one concludes that gauge transformations are
inconsistent with quantum physics. This specific example illustrates the
general argument written in the paragraph that begins below
$(\!\!~\ref{eq:GAUGEQM})$.

\vglue 0.66666in
\noindent
{\bf 5. Tentative Consequences}
\vglue 0.33333in

In this Section several consequences that may result from the foregoing
analysis are presented. This list probably does not exhaust all possibilities.

\begin{itemize}
\item[{1.}] All applications of gauge functions continue to hold within MLE.
Indeed, The equations of motion of MLE - Maxwell equations and the
Lorentz law of force - depend on electromagnetic fields. These fields are
not affected by {\em any} gauge transformation. Hence, all results
derived from these equations remain intact.

As an illustration, let us examine the dimensional and the covariance
arguments discussed in points 1 and 2 of Section 3. Within the
scope of MLE, a gauge transformation adds a zero to the fields. Now a
zero is consistent with {\em all} dimensions and with {\em all}
tensorial quantities. Hence, within MLE, a gauge transformation is
acceptable.

Another example is the solution of Maxwell equations
obtained from an application of the the Green
function of the d'Alembertian (see [1], p. 117; [2], pp. 220, 549).
As is well known from
the theory of differential equations, a solution of a linear homogeneous
equation can be added to a specific solution of the corresponding
inhomogeneous equation. Hence,
gauge transformations remain an important tool for finding a solution
to Maxwell equations.

\item[{2.}] One possibility that may hold for VE is that all operations
and all restrictions associated with gauge transformations
will continue to hold. In this
case, the role of the present paper is to provide a stimulus for an
analysis that will substantiate the freedom of gauge transformations
in VE.
{\em This assignment must settle all gauge related problems derived
above}. Even in this case, MLE and VE are not exactly equivalent because
VE needs the (yet unknown) theoretical structure mentioned in this item
whereas MLE does not need it.

\item[{3.}] Another possibility is that all operations that use
gauge transformations in VE
will continue to hold but gauge invariance will stop to be a mandatory
relation for the acceptability of
electrodynamic expressions. This case may be regarded as
the minimal theoretical change that emerges from
the present work. (As a matter of fact,
this possibility was the initial motivation that has led the Author
to carry out the present research.)

\item[{4.}] A more profound scenario that may result from this work
is that some or all operations which are based on gauge transformations
will be forbidden within VE. In this case electromagnetic relations
which are derived today from gauge related procedures
should be based on other kinds of proofs. In particular, electromagnetic
relations that have been confirmed in experiments are expected to be
proved successfully by other methods.
\end{itemize}

   The alternative scenarios described above illustrate the nature of
this work. It is not intended to present a comprehensive solution of
a physical problem but to draw the attention of the physical community to
a problem which deserves a further analysis.

\vglue 0.66666in
\noindent
{\bf 6. Conclusions}
\vglue 0.33333in

The foregoing results indicate the difference between an electrodynamic
theory where Maxwell equations and the Lorentz law of force are regarded
as the theory's cornerstones and an electrodynamic
theory based on the variational
principle together with
causality and space-time homogeneity. Indeed, if Maxwell
equations and the Lorentz law of force are the theory's cornerstone then
it is very well known that one is free to
define the gauge function $\Phi(x^\mu )$ of $(\!\!~\ref{eq:GAUGE})$
(see [1], pp. 52-53; [2], pp. 220-223). On the other hand,
this work proves that gauge transformations are inconsistent with
electrodynamics based on the variational principle.
In particular, all terms of the Lagrangian density must be Lorentz
scalars having the dimension $[L^{-4}]$. The discussion presented
in this work explains why the variational principle requires the
usage of the Lienard-Wiechert 4-potentials as a unique expression.
For this reason, one concludes
that the two approaches are {\em not equivalent.} It is also proved
that gauge transformations are forbidden in quantum physics.

The outcome of this work does not negate the well known gauge invariance
of the Lagrangian density. Indeed, in the Dirac Lagrangian density
$(\!\!~\ref{eq:DIRACLD})$, the two parts of the gauge transformation
$(\!\!~\ref{eq:GAUGEQM})$ cancel each other. (Hence, the action, the
associated phase and the interference pattern are formally unaffected by a
gauge transformation.) On the other hand, other problems emerge. Thus,
the Dirac Hamiltonian $(\!\!~\ref{eq:DIRACH})$
does not contain the time-derivative of the
gauge transformed wave function $(\!\!~\ref{eq:GAUGEQM})$. Therefore,
one term has no counterpart and the Hamiltonian varies. This
conclusion explains why the Lagrangian density $(\!\!~\ref{eq:DIRACLD})$
is invariant under the gauge transformation $(\!\!~\ref{eq:GAUGEQM})$
whereas the corresponding Hamiltonian is {\em not} invariant under it.
The specific
example discussed above examines a free motionless charged particle.
An application of a gauge transformation casts its Hamiltonian into a
time-dependent expression. This is unacceptable because energy of
a free particle should be a constant of the motion and its Hamiltonian
should be time-independent. Another problem arises in quantum mechanics
because the gauge function $\Phi (x^\mu )$ appears as an exponential factor
of the particle's wavefunction. Hence,
as explained above, it must be a constant and the associated gauge 4-vector
vanishes identically.

   This work aims to examine restrictions imposed on gauge transformations
of electrodynamic systems. It introduces the examination of MLE and VE as
two theories which may be different. The results justify this distinction
because it is proved that any gauge transformation is acceptable within
MLE whereas VE requires a unique 4-potentials.

\newpage
References:
\begin{itemize}
\item[{*}] Email: elic@tauphy.tau.ac.il  \\
\hspace{0.5cm}
           Internet site: http://www-nuclear.tau.ac.il/$\sim $elic
\item[{[1]}] Landau L D and Lifshitz E M, 2005 {\em The Classical
Theory of Fields} (Amsterdam: Elsevier)
\item[{[2]}] Jackson, J D 1975 {\em Classical Electrodynamics}
(New York: John Wiley)
\item[{[3]}] Bjorken J D and Drell S D 1965 {\em Relativistic Quantum
Fields} (New York: McGraw-Hill)
\item[{[4]}] Peskin M E and Schroeder D V 1995 {\em An Introduction to
Quantum Field Theory} (Reading, Mass.: Addison-Wesley)
\item[{[5]}] Landau L D and Lifshitz E M 1960 {\em Mechanics}
(Oxford: Pergamon)
\item[{[6]}] Schiff L I 1955 {\em Quantum Mechanics} (New York: McGraw-Hill)
\item[{[7]}] Bjorken J D and Drell S D 1964 {\em Relativistic Quantum
Mechanics} (New York: McGraw-Hill)

\end{itemize}

\end{document}